\documentclass[12pt]{article}
\usepackage{epsfig}
\usepackage{graphicx}
\usepackage{amssymb}
\usepackage{amsmath}
\usepackage{eqnarray,amsmath}
\usepackage{amssymb}
\usepackage{amsmath}
\usepackage{eqnarray,amsmath}
\begin{document}
\newcommand{\pa}{\partial}
\newcommand{\eps}{\varepsilon}
\newcommand{\lam}{\lambda}
\newcommand{\imag}{\jmath\,}
\newcommand{\Lam}{\Lambda}
\newcommand{\om}{\omega}
\newcommand{\Om}{\Omega}
\newcommand{\Th}{\Theta}
\newcommand{\beq}{\begin{equation}}
\newcommand{\eeq}{\end{equation}}
\newcommand{\sech}{\textrm{sech}}
\newcommand{\qt}{{\bf 1}}
\newcommand{\qi}{{\bf i}}
\newcommand{\qj}{{\bf j}}
\newcommand{\qk}{{\bf k}}
\newcommand{\erf}{{\textrm{erf}}}
\newcommand{\hot}{\textrm{ h.o.t. }} 
\newcommand{\hft}{\textrm{ h.f.t.}} 
\newcommand{\sgn}{\textrm{ sgn}} 
\newcommand{\sinc}{\textrm{ sinc}}
\newcommand{\cosc}{\textrm{ cosc}}
\newcommand{\fbf}{{\bf f}
}

\title{A Frequency-Phase Potential for a Forced STNO Network: an Example of Evoked Memory
\footnote{ Supported in part by Neurocirc LLC. I thank Eugene Izhikevich for commenting on an earlier version of this paper.}
\author{Frank Hoppensteadt\\
Courant Institute of Mathematical Sciences\\
New York University}
}
\date{}

\maketitle

\begin{abstract} 
The network studied here is based on a standard model in physics, but it appears in various applications ranging from spintronics to neuroscience. When the network is forced by an external signal common to all its elements, there are shown to be two potential (gradient) functions: One for amplitudes and one for phases. But the phase potential disappears when the forcing is removed. The phase potential describes the distribution of in-phase/anti-phase oscillations in the network, as well as resonances in the form of phase locking. A valley in a potential surface corresponds to memory that may be accessed by associative recall. The two potentials derived here exhibit two different forms of memory:  structural  memory (time domain memory) that is sustained in the free problem, and evoked memory (frequency domain memory)  that is sustained by the phase potential, only appearing when the system is illuminated by common external forcing. The common forcing organizes the network into those elements that are locked to forcing frequencies and other elements that may form secluded sub-networks.  The secluded networks may perform independent operations such as pattern recognition and logic computations. Various control methods for shaping the network's outputs are demonstrated.
\end{abstract}

In the terminology used here, an oscillation is a signal that is characterized by amplitude, frequency, and phase deviation (timing), and an oscillator is any device that may produce such signals. An oscillatory neural network (ONN) is a network of oscillators designed to achieve some objective, such as learning, storing and retrieving memory, computing, controlling subsystems, or switching. The oscillator network studied here is based on an electromagnetic oscillator called a spin torque nano oscillator (STNO). The space scale of an STNO is 100 or fewer nanometers and one operates at GHz frequencies. Standard methods are used to describe the response of the oscillators when they are connected to others and when the array is exposed to a common oscillating background. The responses are described in terms of potential (gradient) functions for amplitudes and phases, where valleys in the potential manifolds  correspond to
stable patterns of electromagnetic oscillations.

Neuromorphic aspects of this work are partly guided by the following observations from neuroscience: Neurons and networks of them may have oscillatory behavior \cite{buzsaki}; neural tissue has stability properties, including phase and frequency locking \cite{guttman}; EEGs suggest that particular background frequency bands may be important in brain functions \cite{nunez}; resonance phenomena are present in neural activity \cite{emires,burst}; and phase coding has been identified in the brain \cite{emires,okeefe,singer,winfree}. There are also hardware considerations: A single oscillator may efficiently replace a larger network of off-on devices \cite{grol1, grol2}; and, there are well-developed technologies for frequency-based computation, control, signal processing, phase detection, and information flow.

An interesting question in brain science that partly motivates this work is: How might interactions between background fields, whatever their source, and networks of electrically active cells function? The specific network of spin-torque nano-oscillators studied here gives some insight to the functionality of such a structure. A common background signal is shown to enable a network to process and classify information, create and sustain memory, and control the flow of information using frequency based operations. This is an extension of work in \cite{dynconn, actpat, FCH}. This is not a study of the brain. 

The novelty here is deriving a frequency-phase landscape for an STNO network and simplifying its analysis by introducing a phase factor that unfolds singularities in the STNO model.

The STNO model is described first, then it is converted to phase-amplitude coordinates. The amplitude potential is derived, and then the system is further reduced to a flow on a torus where the phase potential is derived. Approximations to the system's states are derived, and these are used to estimate the power spectrum density and to describe various controls for bursting, self-organization, and hysteresis.  Simulations of the system illustrate  its phase potential. Next, a uniform forcing is applied to a weakly connected network of STNOs, and various patterns of phase locking and in-phase/anti-phase responses are described. When the external forcing is replaced by the combined output of the oscillators, referred to here as global feedback, then the phase deviations between oscillators are shown to be synchronized and intervals of phase locking appear. Memory, information flow, self-organization, computation, perturbations by random noise, and continuum models are discussed.

\section{Reduced LLGS model.}
The STNO model studied here is derived from the  Landau-Lifshitz-Gilbert-Slonczewski model  (LLGS) \cite{slavin, macia}. The model describes activity in a thin layer of material that contains magnetic elements. A magnetic element has a magnetization vector in three dimensions, and a complex number $z$ is used to represent the projection of the magnetization vector onto the layer. The oscillator is at rest when $z=0$, meaning that the vector is orthogonal to the layer. According to Maxwell's equations an electric current in the direction of the magnetization may apply torque to the vector causing it to precess. Oscillations in magnetization are useful in a number of applications including computation and memory \cite{macia}.

The model is \cite{FCH,slavin,macia,wcnn}
\beq\label{eq:z}
\dot z = b\frac{z}{|z|}+\lam z-|z|^2z+\imag g(|z|)) z+ \eps f(t)
\eeq
where $\dot z = dz/dt$, $\imag=\sqrt{-1}$, the parameters $b \textrm{ and } \lam$ represent parametric forcing, the frequency modulation term $g$ is a smooth, real-valued function,
and external forcing is represented by $f$, that has effect when the system is otherwise at rest ($z=0$ here).

\textbf{The phase factor, $bz/|z|$.} Including the phase factor $b z/|z|$, which does not appear in  \cite{FCH,slavin,macia,wcnn},  is based on the following reasoning: The equilibrium amplitudes ($R$) with $b=0, \eps=0$, are determined by solving the equation
$$
0 = \lam R - R^3
$$
that has a singularity at $\lam=0, R=0$. There are three real solutions of this equation (counting multiplicity), $R = 0, \sqrt{\lam_+},\, e^{\imag\pi}\sqrt{\lam_+}$, where $\lam_+=(\lam+|\lam|)/2$ is the positive part of $\lam$. These solutions may be designated as zero, spin up, and spin down, respectively, to acknowledge their signs. The universal unfolding of this singularity is
$$
0=b+\lam R - R^3
$$
where $b$ is the unfolding parameter.  When $b>0$, there is a unique positive root of this equation, denoted here by 
$$
R=R^*(b,\lam).
$$ 
It follows that $R^*>\sqrt{\lam_+}$ and $\lim_{b\to +0} R^*(b,\lam)=\sqrt{\lam_+}$. There are two negative real roots that appear through a fold bifurcation when $\lam$ is sufficiently large; those will be considered elsewhere. The phase factor simplifies calculations since $1/R$ often appears in the course of analysis, and $R=0$ cannot happen if $b\ne 0$. However, the phase factor does introduce into equation (\ref{eq:z}) an essential singularity at $z=0$,  which is repelling. 

The phase factor $b z/|z|$ preserves rotational symmetry in the free problem ($f\equiv 0$): If $z$ is a solution, then so is $ze^{\imag\nu}$ for any real number $\nu$.  In particular,  both $z$ and its anti-phase twin $ze^{\imag\pi}=-z$ are solutions of the free problem. A calculation (not presented here) shows that 
$$
z(t,b)-z(t,0)=O(b^{1/3})+O(\eps) 
$$
uniformly in $t\ge t^*(b)>0$, where $t^*(b)$ goes to zero as $b\to +0$, accounts for an initial transient \cite{cims22}. No physical argument is presented here for including the phase factor. 

\textbf{External forcing}. The external (or background) forcing $f$ comprises $M$ modes of oscillation:
\beq\label{eq:f}
f(t)=\sum_{m=1}^M C_m e^{\imag(\Om_m t + \Psi_m)}
\eeq
having (real) amplitudes $C_m$, frequencies $\Om_m$, and phase deviations $\Psi_m$.  Without loss of generality, the frequencies may be relabelled so $\Om_1<\Om_2<\dots <\Om_M$. These frequencies may be incommensurable, in which case $f$ is quasi-periodic. Also, $C_m\ge 0, \textrm{ for } m\in\overline{1,M}$,\footnote{The notation $\overline{1,M}=\{1,\dots,M\}$ is used.} since their signs may be set by adjusting $\Psi_m$ by adding $\pi$ or not. The background field is assumed to be weak relative to the dynamics of the oscillator, as represented by the small positive parameter $\eps$ in (\ref{eq:z}).

\textbf{Phase-amplitude coordinates.} Changing $z$ to polar coordinates, $z(t)=R(t)\exp(\imag\theta(t))$, takes equation (\ref{eq:z}) into 
$$
\dot R + \imag R\,\dot\theta= \left[b+(\lam+\imag g(R)) R-R^3\right] + \eps\sum_{m=1}^M C_m e^{\imag(\Om_m t+\Psi_m-\theta)},
$$
after dividing both sides by $\exp(\imag\theta)$. Equating real and imaginary parts gives a system of two equations, one for the amplitude $R$ 
\beq\label{eq:R}
\dot R = (b+\lam R-R^3) + \eps\sum_{m=1}^M C_m\cos(\Om_m t+\Psi_m-\theta),
\eeq
and one for the phase $\theta$
\beq\label{eq:theta}
R\,\dot\theta =  g(R)R+ \eps\sum_{m=1}^M C_m\sin(\Om_m t+\Psi_m-\theta).
\eeq

\section{Gradient Analysis of (\ref{eq:z})} \label{s:results}

Let $z$ be determined from equation (\ref{eq:z}) where $f$ is given in (\ref{eq:f}). Following are some facts about this system.

\subsection{Amplitude potential.} The amplitude potential function used here is defined by
\beq\label{eq:V}
V(R; b,\lam)=\left(\frac{R^4}4-\lam \frac{R^2}2 -bR\right).
\eeq
With this, equation (\ref{eq:R}) may be rewritten as
$$
\dot R = -\frac{\pa V}{\pa R}(R; b,\lam)+ \eps\sum_{m=1}^M C_m\cos(\Om_m t+\Psi_m-\theta),
$$
where the last term approaches zero as $\eps\to 0$ uniformly for $0\le t < \infty$.
The potential function $V(R; b,\lam)$ describes a landscape of amplitudes indexed by $b$ and $\lam$. 

The isolated minima of  $V$ define equilibria for the system when $\eps=0$, and they are stable under persistent disturbances. Stable under persistent disturbances here means that small perturbations of a gradient  system $\dot x = -\nabla F(x)$, say $\dot y = -\nabla F(y)+\eps g(t,y)$, preserve stable behavior near minima of $F$. That is, if $x^*$ is an isolated minimum of $F$ and $y(0)$ is near $x^*$, then,  under natural conditions\footnote{For example, the perturbation $g(t,y)$ is bounded, $L^1$ as a function of $t$, and differentiable as a function of $y$.} on $g$,  $y(t)=x^*+$ error, where the error goes to 0 as $\eps\to 0$ uniformly for all $0\le t <\infty$.   Stability under persistent disturbances is justified by Lyapunov's stability theory since the potential defines a Lyapunov function at each of its minima \cite{malkin, ascs}.

\subsection{Frequency-Phase potential.} 

The convergence of $R$ to $R^*$ occurs first among the hierarchy of time scales in this problem as $\eps\to 0$ when $b>0$, and in most of the calculations done here we assume that $R$ is at its asymptotic value $R=R^*(b,\lam)$. The remainder of this paper is based on (\ref{eq:z}) with $g(|z|)$ replaced by the constant $\om^*=g(R^*(b,\lam))$, since the difference $R(t)-R^*=O(\exp((\lam-3R^*) t/\eps)$.

The surface $R=R^*$ defines a torus that is indexed by $\theta$ and $\vec\Om\, t$ where $\vec\Om$ is the vector of forcing frequencies,  and the model studied in the remainder of this section is
\beq\label{eq:torus}
\dot\theta = \om^* + \eps\sum_{m=1}^M \hat C_m \sin(\Om_m t +\Psi_m-\theta)
\eeq
where $\hat C_m=C_m/R^*$. This equation represents the reduction of (\ref{eq:z}) to a differential equation on an $M+1$-dimensional torus.

The differences in forcing frequencies are assumed to satisfy $|\Om_k-\Om_l |\gg\eps$ for all $k\ne l$. Setting $\theta=\om^* t + \phi$ in equation (\ref{eq:torus}) gives an equivalent equation, but now one for the phase deviation $\phi$. Equation (\ref{eq:torus}) becomes
\beq\label{eq:phi}
\dot\phi= \eps \sum_{m=1}^M \hat C_m\sin((\Om_m-\om^*) t+\Psi_m-\phi).
\eeq

Interest here is in behavior over a finite interval $0\le t\le T/\eps$  for some fixed time $T$; that is, for the slow time variable $s=\eps t$ ranging over $0\le s\le T$. The method of averaging may be applied to (\ref{eq:phi}) as follows. Define the averaging operator acting on an integrable function $F(t)$ to be
$$
\Big<F\Big>\equiv \frac{\eps}{T}\int_0^{T/\eps} F(t')\,dt'.
$$
Applying this operator to (\ref{eq:phi}) while viewing $\phi=\phi(s)$ as being an independent function of the slow time $s$, gives
$$
\Big<\frac{d\phi}{ds}\Big>= \sum_{m=1}^M \hat C_m\Big<\sin((\Om_m-\om^*) t+\Psi_m-\phi(s))\Big>.
$$
This average evaluates to
\beq\label{eq:barphi}
\frac{d\bar\phi}{ds}=\sum_{m=1}^M \hat C_m\left(\frac{\cos(\bar\phi-\Psi_m)-\cos(\bar\phi-\Psi_m-\nu_m)}{\nu_m}\right)
\eeq
where $\bar\phi$ replaces $\phi$ in the average and $\nu_m=(\Om_m-\om^*)T/\eps$. 

This equation may be rewritten as
$$
\frac{d\bar\phi}{ds}=-\frac{\pa W}{\pa\bar\phi}(\bar\phi,\om^*,\eps)
$$
where
\beq\label{eq:Wshort}
W(\bar\phi,\om^*,\eps)=-\sum_{m=1}^M \hat C_m\frac{\sin(\bar\phi-\Psi_m)-\sin(\bar\phi-\Psi_m-\nu_m)}{\nu_m}.
\eeq
An equivalent form of $W$ that is useful for analysis and simulation is\footnote{The unnormalized cardinal functions used here are $\sinc \,x \equiv \sin x/x$, and $\cosc\,x\equiv (1-\cos x)/x$. Note that $\sinc(0)=1, \cosc(0)=0$, and $\cosc(x)=(x/2) \sinc(x/2)^2$.}
\beq\label{eq:Wsinc}
W=-\sum_{m=1}^M \hat C_m(\cosc(\nu_m)\sin(\bar\phi-\Psi_m)+\sinc(\nu_m)\cos(\bar\phi-\Psi_m)).
\eeq
The averaging method shows that $|\phi(t)-\bar\phi(\eps t)|=O(\eps)$ uniformly for $0\le t \le T/\eps$ \cite{ascs}. 
The isolated minima of $W$ are equilibria for $\bar\phi$, and these are stable under persistent disturbances. Note for later use that
\beq\label{eq:Wphi}
\frac{\pa W}{\pa\om}\Big|_{\om=\om^*}=\frac{T}{\eps}\sum_{m=1}^M \hat C_m(\cosc^*(\nu_m)\sin(\bar\phi-\Psi_m)+\sinc^*(\nu_m)\cos(\bar\phi-\Psi_m))
\eeq
where $\sinc^*(x)=d\sinc(x)/dx$ and $\cosc^*(x)=d\cosc(x)/dx$.

\subsection{Phase and Frequency Locking.} 
How wide are the valleys in $W$ near its minima? Define
$$
X_K(\om^*)=\frac{(\Om_K-\om^*) R^*}{\eps C_K}.
$$
It is shown in Appendix 1 that if the phase locking condition
\beq\label{eq:PLcondition}
|X_k|<1
\eeq
holds, then
$$
\theta \approx \Om_K t +\phi^*_K
$$
where $\phi^*_K=\Psi_K- \arcsin X_K$, and the output frequency is $\Om_K$. The error in this approximation is shown in Appendix 1.
 
In summary, if (\ref{eq:PLcondition}) holds, then the amplitude approaches $R^*$, the frequency is locked to $\Om_K$, and the phase deviation $\phi_K^*=\Psi_K-\arcsin X_K$ is stable. In this situation
$$
z = R^* e^{\imag(\Om_K t + \phi_K^*)}+ \textrm{remainder}
$$
where $R^*$ and $\phi_K^*$ are at or near stable minima of  $V$ and $W$, respectively, and the remainder is small for large $TR^*/\eps$ and for small $(\Om_K-\om^*)$.

Estimates of the width of wells in $W$ may be derived from (\ref{eq:PLcondition}). They are the frequency intervals
$$
\Om_m-\frac{\eps C_m}{R^*}<\om<\Om_m+\frac{\eps C_m}{R^*}
$$
for $m\in\overline{1,M}$; this interval lies within a well of $W$ located near $\Om_m$ provided $|\eps C_m/R^*|$ is sufficiently small. This shows to what extent the power in the forcing signal ($|C_m|^2$) controls the size of phase locking intervals.

Rather than a detailed analysis of the extrema of $W$, 
their nature is demonstrated in the  limit $T/\eps\to\infty$, where $W$ in (\ref{eq:Wsinc}) becomes
\beq\label{eq:Winfty}
W_\infty(\bar\phi,\om)=-\sum_{m=1}^M \hat C_m\,\delta(\om-\Om_m)\,\cos(\bar\phi-\Psi_m).
\eeq 
where $\delta$ is Kronecker's function: $\delta(0)=1$ but $\delta(x)=0$ for $x\ne 0$. The function $W_\infty$ has minima at $\om=\Om_m, \bar\phi=\Psi_m$ for $m\in\overline{1,M}$, and their depth is proportional to $\hat C_m\equiv C_m/R^*$. By continuity, $W$ will have minima near these values, as shown by direct calculation that is not done here. 

\textbf{Power spectrum}. The power spectrum density (PSD) of $z$ is derived in Appendix 2. The result is that
\beq\label{eq:psdz}
PSD_z(\om^*)\approx R^{*2}\,\sinc^2((\Om_K-\om^*) T/\eps)+O(\eps/T).
\eeq
for $\om^*\sim\Om_K$. The notation $\om^*\sim\Om_K$ means that the phase locking condition (\ref{eq:PLcondition}) is satisfied.

It is useful to consider also the rotation number for this system
$$
\rho(\om^*)=\lim_{t\to\infty}\frac{\theta(t,\om^*)}{t}
$$
since $\theta$ is accessible in this simple model. This represents the output frequency of the oscillator, and it gives an alternate depiction of the PSD, see Figure 1.

\subsection{Quasistatic parameters.} 

All of the data in (\ref{eq:z}) are candidates for being control variables. The data may be set by external or internal control laws. Three examples of parametric control are given:

BURSTING: The amplitude approximation described above is  $|z|\approx R^*(b,\lam)$. If the parameter $\lam$ is slowly varying, say $\lam=\lam(\mu t)$ where $\mu\ll \eps$, and it executes a cycle increasing through the value $\lam=0$, and back again, then $R^*(b,\lam(\mu t))$ varies from low values near $R=b^{1/3}$ when $\lam<0$ to high values (approximately) $R=\sqrt{\lam(\mu t)}$ when $\lam>0$, and then back to near 0. In that case, the form of $z$ shown in the previous result for $\om$ near $\Om_K$ is
$$
z\approx\sqrt{\lam(\mu t)_+}\,e^{\imag(\Om_K t +\phi_K^*)}.
$$
This will exhibit a burst of activity where the length of a burst is controlled by the  duty cycle of $\sqrt{\lam(\mu t)_+}$ and the frequency within a burst is controlled by $\om$. In biological settings, $\lam$ would typically have fill-and-flush dynamics, for example, exhaustible resources are needed to sustain its activity and once depleted must be replenished, and in electronic circuits by a Schmitt trigger \cite{carhop} for example. 

SELF-ORGANIZATION: If $\om(\mu s)$ is controlled to be slowly varying through its range, then the solution locks onto various forcing frequencies as $\om(\mu s)$ passes through intervals of phase locking. Furthermore, if $W$ is used as a feedback control rule, say
\beq\label{eq:omdot}
\frac{d\bar\om}{ds}=-\mu\frac{\pa W}{\pa\om}(\bar\phi,\bar\om),
\eeq
where the right hand side is given in (\ref{eq:Wphi}), then for $\bar\phi$ near $\Psi_K$ and for $\bar\om$ near $\Om_K$, 
$$
\frac{d\bar\om}{ds}=-\mu\,\hat C_K\Big(\sin(\bar\phi-\Psi_K)/2-2(\Om_K-\bar\om)\cos(\bar\phi-\Psi_K)/3 +\hot\Big)
$$
where $\hot$ denotes higher order terms in $\bar\om-\Om_K$. This equation shows that $\bar\om=\Om_K$ is stable under persistent disturbances, and the frequencies try to organize at the forcing frequencies, depending on their initialization.

HYSTERESIS: If $b$ is controlled by negative feedback from the amplitude, say
$$
\dot b =-\mu R+g(t), \quad\quad\dot R = b+\lam R-R^3,
$$
then differentiating the second equation gives van der Pol's equation
$$
\ddot R +(3R^2-\lam)\dot R +\mu R = g(t).
$$
Therefore, with $b$ being controlled by negative feedback from $R$ (now $b$ may take on negative values), the forced system may exhibit a wide range of behaviors, including negative differential resistance, hysteresis, and chaotic behavior \cite{cims22,flahop}.

\subsection{Simulations.} 

\begin{figure}[htb!] 
\includegraphics[width=3.5in]{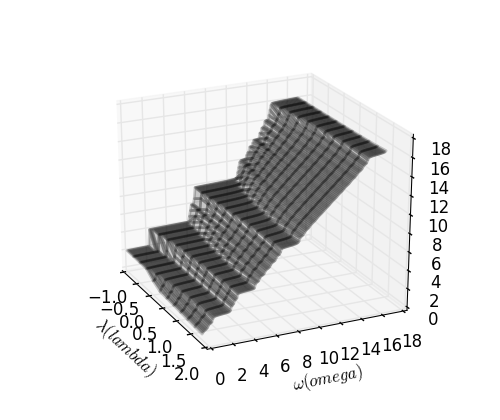}
\caption{Frequency locking: Output frequencies as a function of $\om$ and $\lam$. Projecting plateau regions onto the $\lam\om$-plane define regions where the output frequency is locked, as read off the vertical axis. In this simulation $C_m=10, \Psi_m=0,\Om_m=2^m, m\in\overline{1,4}$, $b=0.5$, $\eps=0.1$, and the output frequency $\rho= \theta(10)/10$. Note that for $\lam<0$, subharmonics appear, as indicated by the finer stair-step structure at $\lam=-1.0$.}
\label{fg:RhoOmLam}
\end{figure}

\begin{figure}[htb!] 
\includegraphics[width=3.5in]{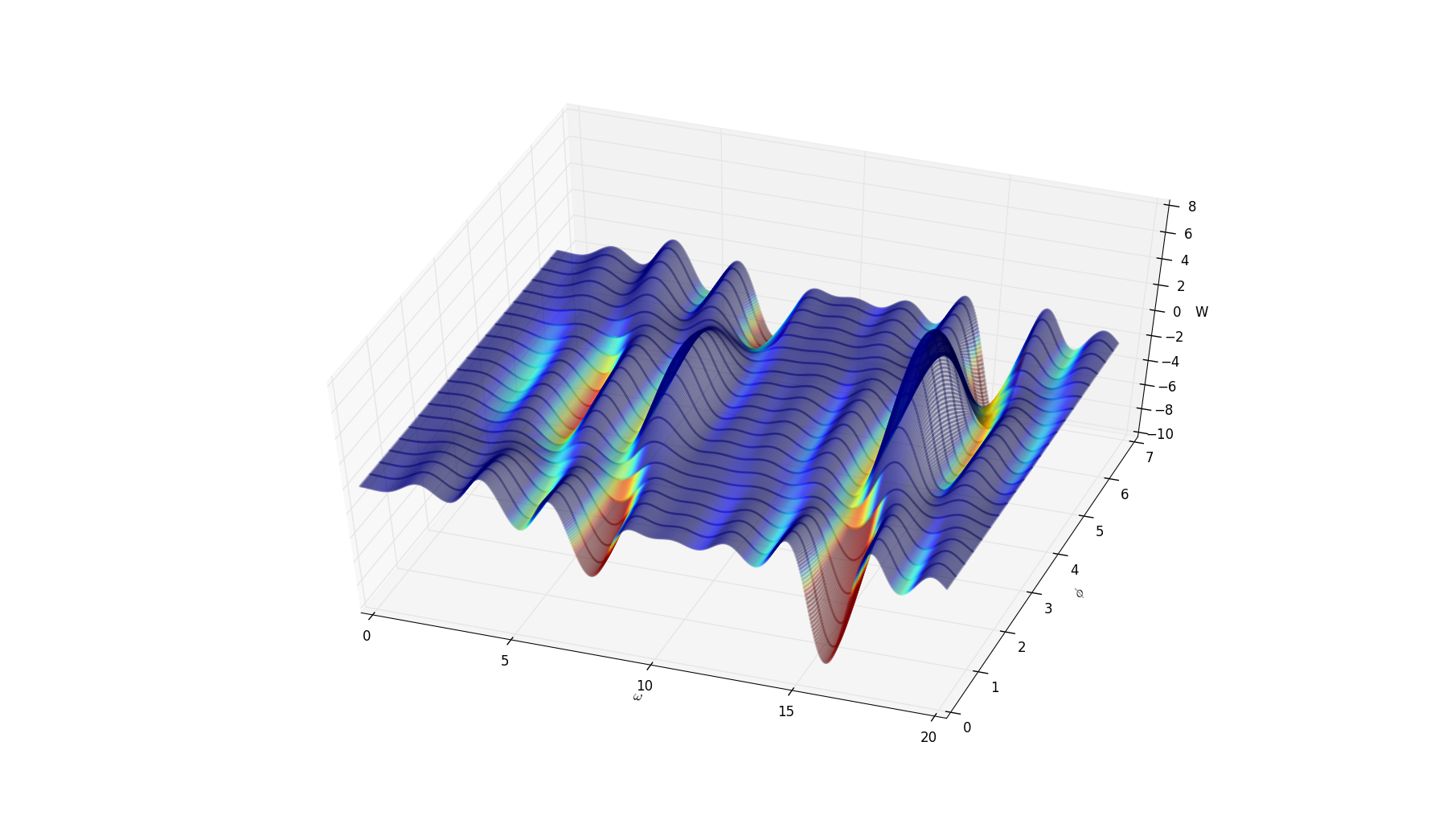}
\caption{The potential function $W$ in (\ref{eq:W4}) is plotted using the same data as in Figure \ref{fg:RhoOmLam} and $T=1$. If $\om$ is near $\Om_k$, then $\phi$ is captured in an energy well corresponding to frequency $\Om_k$. The depth of a  well is related to the power in the corresponding forcing mode ($|C_m|^2$). The width of a well is controlled by the parameter $\eps C_m/R^*$. In this simulation, $\Psi=[\pi, \pi, 0, 0]$, and wells are seen for $\phi$ near $\pi$ for $\om$ near 2 and 4. Wells for $\phi$ near $0\mod 2\pi$ are seen for $\om$ near 8 and 16.}
\label{fg:Womphi}
\end{figure}

Consider equation (\ref{eq:z}) with $M=4$:
\beq\label{eq:eeg}
\dot z = \left(b\frac{z}{|z|} + \lam z - |z|^2z\right)+\imag\om z + \eps\sum_{m=1}^4 C_m e^{\imag(\Om_m t+\Psi_m)}.
\eeq
The gradient potentials for amplitudes and phases in this case are, respectively,
$$
V(R,b,\lam)=R^4/4 - \lam R^2/2 -bR 
$$
and
\beq\label{eq:W4}
W=-\sum_{m=1}^4 \hat C_m\left(\sinc(\nu_m)\cos(\phi-\Psi_m)+\cosc(\nu_m)\sin(\phi-\Psi_m)\right)
\eeq
where $\nu_m=(\Om_m-\om) T/\eps$. The output frequencies are plotted in Figure \ref{fg:RhoOmLam} and $W$ is plotted in Figure \ref{fg:Womphi}.

In Figure~\ref{fg:RhoOmLam} there are plateau regions describing locking to the input frequencies $\vec\Om = (2, 4 , 8, 16)$ for various values of $\om, \lam$. These locking regions correspond to evoked memory in the system when it is illuminated by energy from the background field $f$ \cite{actpat}. Figure \ref{fg:RhoOmLam} shows the changes in the system's response as the parameter $\om$ varies through an interval containing all of the forcing frequencies. The output frequencies that result from $\om$ being in various places are shown in terms of the rotation number $\rho(\om,\lam)=\lim_{t\to\infty}\theta(t)/t$.

The phase locking condition (\ref{eq:PLcondition}) gives an estimate of the domain of attraction of $\phi^*_K$ (in terms of $\om$). Condition (\ref{eq:PLcondition}) is satisfied when $(\Om_K-\om)R^*/(C_K\eps)$ is small and when $C_K$ is large. However, as noted after equation (\ref{eq:xiavg}), the error in that approximation holds for each $R^*>0$, but not uniformly for $R^*>0$. While making $R^*$ near zero may make $X_K$ smaller, the error in  averaging in (\ref{eq:xiavg}) may be large; that notwithstanding, Figure 1 indicates that phase locking strengthens when $\lam<0$ where $R^*\approx 0$. The analysis just completed explains the structure in Figure 2 for $\lam >0$. Additional wrinkles in $W$, those not located at the driving frequencies $\Om_m$, are not spurious minima, since they may correspond to various harmonic responses \cite{flahop} that are not explored further here. The simulation figures may be viewed as being like Bode plots for the phase response ($\phi^*$). The results of the analysis may be plotted  in various other ways depending on the application.

\section{Weakly Connected STNO Networks}

A weakly connected network of $N$ oscillators based on (\ref{eq:z}) is described by the equations
\begin{eqnarray}\label{eq:newnetz}
\dot z_n &=& b_n\frac{z_n}{|z_n|}+(\lam_n +\imag \om_n)z_n -|z_n|^2z_n\\
 &\,&+\,\eps f(t)+\eps^2\sum_{k=1}^N A_{n,k}\,e^{\imag\psi_{n,k}}\,z_k\nonumber
\end{eqnarray}
for $n\in\overline{1,N}$, where $f$ defined in (\ref{eq:f}) represents a forcing that is common to all elements in the array, each $b_n>0$, and, per the earlier discussion, the center frequencies $\om_n$ are constants. The connection matrix is ${\bf A}=(A_{k,l}\exp(\imag\psi_{k,l}))$ for $k,l\in\overline{1,N},$ where $A_{k,l}\ge 0$, $A_{k,k}=0$. Again, a connection's sign may be adjusted by adding to $\psi_{k,l}$ either $\pi$ or 0. The connection matrix ${\bf A}$ describes interactions between network elements, and it may represent electrical or optical connections from one oscillator to another, or torque transfer. The background signal is stronger than the connections between oscillators as reflected by the relative sizes of the small parameters $\eps$ and $\eps^2$.
The analysis is simplified by the fact that to order $\eps$ these oscillators are independent so each component in the array may exhibit self-sustained oscillations, and each has its own  potential function.

Converting to polar coordinates ($z_n=R_ne^{\imag\theta_n}$) and taking 
$R_n=R^*(b_n,\lam_n)$ leads to equations for the phases
\begin{eqnarray}\label{eq:phasenet}
\dot\theta_n &=& \om_n + \eps\sum_{m=1}^M \hat C_{n,m}\sin(\Om_m t + \Psi_m-\theta_n)\\
&\,&+\, \eps^2\sum_{k=1}^N \hat A_{n,k}\sin(\theta_k-\theta_n+\psi_{n,k})\nonumber
\end{eqnarray}
where  $\hat C_{n,m}=C_m/R^*_n$, and $\hat A_{n,k}=A_{n,k}R^*_k/R^*_n$. 
The free problem ($f\equiv 0$) is a weakly connected Kuramoto network
$$
\dot\theta_n = \om_n + \eps^2\sum_{k=1}^N \hat A_{n,k}\sin(\theta_k-\theta_n+\psi_{n,k})
$$
that has been extensively studied (e.g., \cite{wcnn}). This not a gradient system unless ${\bf A}$ satisfies some restrictive conditions. 

As before, we focus on the phase deviations $\phi_n=\theta_n-\om_n t$:
\begin{eqnarray}\label{eq:phinet}
\dot\phi_n &=& \eps\sum_{m=1}^M \hat C_{n,m}\sin((\Om_m-\om_n) t + \Psi_m-\phi_n)\\
&+&\eps^2\sum_{k=1}^N \hat A_{n,k}\sin((\om_k-\om_n)t+(\phi_k-\phi_n)+\psi_{n,k})\nonumber
\end{eqnarray}

Averaging system (\ref{eq:phinet}) over $T$, gives
$$
\dot{\bar\phi}_n=-\eps\frac{\pa W}{\pa\bar\phi_n}(\bar\phi_n,\om_n,\eps)+O(\eps^2)
$$
where $W$ is defined in (\ref{eq:Wsinc}).

Denote by $S_K$ the set of center frequencies $\om_n$ with $\om_n \sim \Om_K$ for  each $K\in\overline{1,M}$. Then the calculation in Appendix 1 shows that
$$
\theta_n\approx \Om_K t +\phi^*_K.
$$
For $n$ such that $\om_n\notin S\equiv\cup_{k\in\overline{1,M}} S_k$
$$
\frac{d\bar\phi_n}{ds}=\eps\sum_{k=1,\om_k\sim\om_n}^N \hat A_{n,k}\sin(\bar\phi_k-\bar\phi_n+\psi_{n,k}).
$$
This is referred to as being a network secluded from $f$. It is further structured by those frequencies that match others and those that don't. The latter oscillators wander, the former create coherent subnetworks called affinity groups.
This shows that the forcing divides the network into those oscillators that are locked to a forcing mode and those that are not, but the latter may form affinity groups of oscillators that are  Kuramoto networks having identical center frequencies.

Note that the function
$$
\tilde W(\vec\phi,\vec\om)=\sum_{n=1}^NW(\phi_n,\om_n,\eps)
$$
where $W$ is given in (\ref{eq:Wsinc}) defines a potential function for the averaged network to leading order in $\eps$.

\subsection{Global Feedback.}
Electroencephalograms (EEGs) describe oscillatory electromagnetic activity in a brain.  The signals being measured are believed to be generated by sources within the neocortex \cite{nunez}. One view is that EEGs represent the composite activity of a population of cells, which in turn creates an oscillatory environment in which the network operates. 

To study how the STNO network operates in this case, we replace $f+\eps {\cal A}z$ by $\frac 1N \sum z_n$:
\beq\label{eq:globalfeed}
\dot z_n=b\frac{z_n}{|z_n|}+(\lam_n+\imag \om_n) z_n-|z_n|^2z_n +\frac \eps N \sum_{k=1}^N z_k.
\eeq
No assumption is made about the physical positioning of the oscillators; they may be contiguous or widely dispersed. The only thing they share is the input from their aggregate output.

Converting to polar coordinates, setting $\phi_n=\theta-\om_n t$, and averaging as for (\ref{eq:phi}) gives
\beq\label{eq:globalphibar}
\dot{\bar\phi}_n=\eps\frac 1N \sum_{n=1}^N \frac{R^*_k}{R^*_n}c_{n,k}\sin(\bar\phi_k-\bar\phi_n)+O(\eps^2)
\eeq
where
$$
c_{n,k}=\sinc\left(\frac{T(\om_k-\om_n)}{\eps}\right).
$$
The coefficients may be symmetric since $c_{k,l}=c_{l,k}$ for all $k,l\in\overline{1,N}$, but in general $R_k/R_l\ne R_l/R_k$. 

\begin{figure}[h!] 
\includegraphics[width=3.5in]{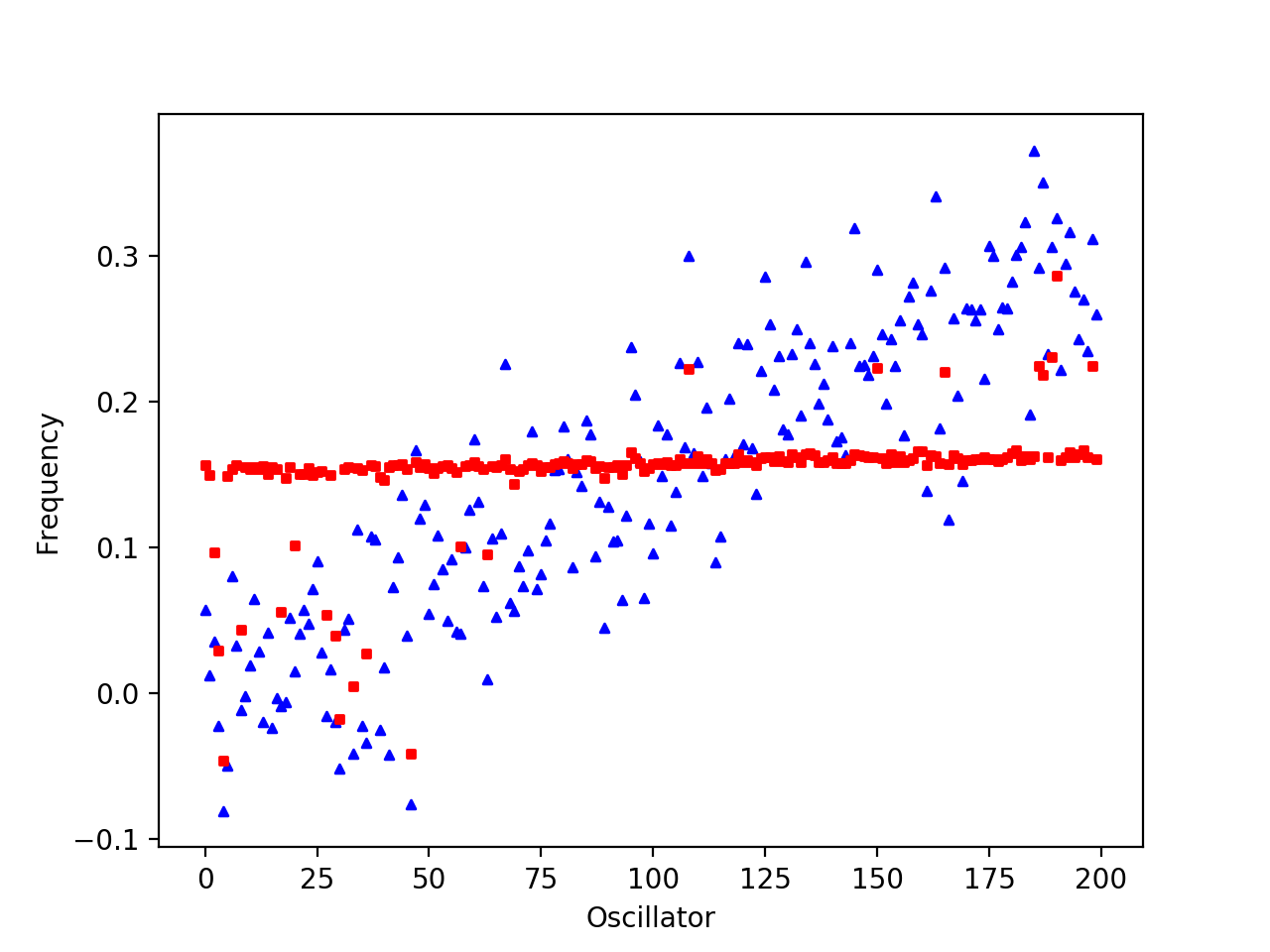}
\caption{This simulation plots the input and output frequencies of 200 oscillators when the center frequencies form a gradient $\om_n=\pi n/2000 + \eta_n$  for   $\eta_n\in{\cal N}(0,0.05)$, $n\in\overline{1,200},$ and the amplitudes are randomly chosen: $R_n^*\in{\cal U}$, which is the uniform distribution on $(0,1]$.  Input frequencies ($\om_n$, triangle markers) and output frequencies ($\rho_n$, square markers) are plotted for 200 oscillators.  Equation (\ref{eq:globalphibar}) is solved for $0\le t\le 130$, $\eps=0.5, \lam=1.0,$ and $T=1.0$. This demonstrates that global feedback stabilizes the output frequency of the population. In most cases, the output frequencies converge to the average of the center frequencies, which is $\bar\om=0.1497$, and the EEG of this system, not shown here, has a large spike at $\rho=\bar\om$. The deviations of $\rho_n$ from $\bar\om$ are largely due to random variations in the ratios $R^*_n/R^*_k$.}
\label{fg:globalrho6}
\end{figure}

If all $R_n^*=R^*$ are identical, a standard argument, e.g. Theorem 11.4 \cite{wcnn}, shows that for each $n\in\overline{1,N}$ the phase deviation $\bar\phi_n$ has a limit, say $\bar\phi_n\to\bar\phi_n^*$. Suppose the center frequencies have a common random distribution, say for each $n\in\overline{1,N}$, $\om_n\in{\cal N}(\Om,\sigma^2)$, which is a normal gaussian random variable having mean $\Om$ and variance $\sigma^2$. Then the coefficients 
$$
c_{n,k}=\sinc\left(\frac{T(\om_k-\om_n)}{\eps}\right)
$$
form a symmetric matrix, and the arguments of $\sinc$ are identically distributed random variables having a normal distribution. Averaging over ${\cal N}$ gives the averaged coefficients
$$
\bar c=\int_{-\infty}^\infty \sinc(\eta)\,dw(\eta)
$$
where  $w$ is the probability measure for ${\cal N}$ \cite{skor}. Setting $\xi_n = (\Om-\om_n) t -\bar\phi_n$ in (\ref{eq:globalphibar}) and averaging over $\cal N$ and over $T/\eps$ gives
$$
\dot\xi_n=\eps\frac 1N \,\bar c\,\sum_{k=1}^N\sin(\xi_k-\xi_n).
$$
A standard argument shows that the variables $\xi_n$ have limits
$$
\lim_{t\to\infty}\xi_n(t)=\chi_n^*,
$$
and $\theta_n\approx \bar\om\, t + \chi^*_n$, where $\bar\om$ is the average of the set $\{\om_n\}$. The approximation is in a probabilistic sense not made precise here, except to say that the approximation degrades with increasing variance $T^2\sigma^2/\eps^2$. In this case the global feedback synchronizes the oscillators with the phase deviation distribution $\chi^*$ and the output frequencies are $\bar\om$. A simulation of this is shown in Figure \ref{fg:globalrho6} where both the center frequencies and the amplitudes $R^*_n$ are randomly chosen.

Figure \ref{fg:globalrho6} shows that global feedback within the network may stabilize it at a common frequency, and small random perturbations of the data do not significantly impact this result. If there are additional oscillators that have their center frequencies similarly clustered but sufficiently separated, the feedback will stabilize the cluster to a single frequency and the PSD of $z$ will have spikes at the locked frequencies. No assumption is made about the spatial location of oscillators; but, the results here show coherence in the frequency domain when the composite signal feedbacks to the network.

There are many more interesting problems here for stochastic analysis, but we consider only this case to illustrate frequency and phase-locking features of the global feedback network. In particular, global feedback may synchronize the population of phase deviations and lock the oscillator frequencies. 

\section{Discussion}

The aims of this paper were to demonstrate how the network (\ref{eq:newnetz}) can create and sustain memory, process and classify information, and perform computations.

\textbf{Memory:} The valleys in a potential may be used as memory. These may be accessed by associative recall since valleys correspond to basins of attraction of their minima. That is, if the system is initialized in a valley, it will reliably converge toward a corresponding minimum of the potential. The analysis here identified  the phase potential function, located its basins of attraction, and estimated their scope. 

There are two kinds of memory in system (\ref{eq:z}), one called here structural, or time domain,  memory,  and the other evoked, or frequency domain, memory. The function $V$ with $b>0, \lam>0$, has a single positive minimum at $R^* \approx\sqrt{\lam}$, although the analysis here supports $V$ having more wells.
Evoked memory is described by the phase potential. Phase states are stabilized by external forcing, somewhat similar to the up position of a pendulum being stabilized by vibration of its support point, but which becomes unstable when the vibration is removed \cite{actpat}. The phase potential $W$ represents $M$ evoked memory units per oscillator. In the network there are possibly $M\times N$ evoked memory units, depending on the distribution of center frequencies. 

\textbf{Information flow:} The forcing frequencies in $f$ define channels for transferring information using phase deviations.  The oscillators act as phase-locked loops  \cite{vcon} in that their phases converge to forcing phases. If information is encoded in the forcing phases, say $\Psi_m(t)$, as in phase modulation signaling, the network will track these phases in a stable way, and the results may be transmitted by the outputs $\{z_n\}$. The power spectrum density (PSD) of $\{z_n\}$ for $n\in\overline{1,N}$ can be used to describe the channels for communication and their robustness.

\textbf{Computation:} In-phase and anti-phase oscillations in ONN may be used for a variety of computations. Phase deviations may carry digital information in oscillations. For example, the bit `0' corresponds to $\Psi=0$ and the bit `1' to the anti-phase oscillation where $\Psi=\pi$.  $z$ carries this information forward in the various channels established by the forcing frequencies,  and the information may be retrieved from $z$ by correlations \cite{FCH}. 

\textbf{Free system:} If the forcing is not present ($f\equiv 0$), then (\ref{eq:newnetz}) becomes a Kuramoto network, which generally is not a gradient system (e.g., \cite{wcnn}). The potential functions for single oscillators were constructed by straightforward integration of scalar equations. This may not be possible for more general systems. But the analysis here provides guidance when generalizations of gradient systems, such as arrays of local Lyapunov functions or radial basis functions, may be used. 

\textbf{Random noise perturbations:} Stability under persistent disturbances as used here also applies to systems perturbed by small random noise. Large deviation random perturbations may be studied using the theory developed in  \cite{wentzel,skor} for random dynamics in gradient systems. However, with random noise comes the possibility that $R=0$ even if $b>0$, so the solution may be driven through the essential singularity at $z=0$, and so it may experience unknown phase shifts, or cycle slips, which might be near $\pi$ or not.

\textbf{Continuum model:} For a planar array of STNO each magnetic element is described by its magnetization, taken here to be the complex-valued function $z(x,t)$ of time $t$ and a spatial variable $x\in E^2$.
The Landau–Lifshitz–Gilbert–Slonczewski equation (see \cite{slavin,macia})
describes precessional motion of magnetization in a layer, and it accounts for spin-transfer torque from one oscillator to another. The analog of this in our context is the nonlinear Schr\"odinger equation \cite{spinwave}
\beq\label{eq:llgs}
\imag\dot z = (1+\imag \beta) D\nabla^2 z - \imag F(z) + G(|z|) z + \eps f(t).
\eeq
where $D$ accounts for torque transfer between elements, $F(z)=b z/|z|+\lam z - |z|^2 z$, $G$ is a smooth function satisfying $G(0)>0$, and $\beta$ is a damping coefficient. Direct calculation shows that this problem, to leading order in $\eps$, has spin waves of the form
$$
z=R^*e^{\imag(\vec k\cdot x + (1+\imag \beta)D \,|\vec k|^2 t)}e^{\imag\om^*t}
$$
having wave vector $\vec k$ and frequency $\om^*=G(R^*)$.
Spin waves and vortex solutions have been found for this equation when $b=0, \eps = 0$, and excitations in such media may propagate as spin waves that may carry information for computations \cite{FCH,macia,spin2}.  A second centered-difference discretization scheme for (\ref{eq:llgs}) will have the form of (\ref{eq:newnetz}). 

\section{Conclusion}

The response to external forcing by a single STNO with phase factor  was studied first. It was shown that there are potential functions for both the amplitudes and phases. The phase potential was illustrated in the simulations in Figures 1 and 2. The averaged potential function $W_\infty$  gives a useful approximation to $W$, and it provides a convenient description of the  location of energy wells and of the distribution of in-phase/anti-phase behavior when the forcing phase deviations are $0$ or $ \pi$, respectively. If one defines the energy of an oscillator to be proportional to its frequency, the potential function $W$ in (\ref{eq:Wshort}) represents a landscape of energy for the forced oscillator. The analysis here shows how, given the Fourier decomposition of the forcing $f$, there are potential functions, $V$ for amplitude and $W$ for phase deviations for weakly forced networks, as well. The data in $V$, $W$ and $f$ shape the responses of the system, and these data  may be used as control variables to place equilibria and to shape their domains of stability, both in amplitude and phase. These results facilitate using ONN for training and learning, for example by showing how to manipulate outputs to produce  patterns that agree with target outcomes.

\section{Appendix 1: Derivation of phase locking condition}
Consider (\ref{eq:torus}) with $\om^*$ being near to $\Om_K$ for some $K\in\overline{1,M}$. The equation may be rewritten as
$$
\dot\phi=\eps \hat C_K\sin((\Om_K-\om^*) t+\Psi_K-\phi)+ \eps \sum_{m=1,m\ne K}^M \hat C_m\sin((\Om_m-\om^*) t+\Psi_m-\phi).
$$
Let a new variable $\xi$ be defined by $\xi=(\Om_K-\om^*) t+\Psi_K-\phi$. 
This gives on the slow time scale
\begin{eqnarray}\label{eq:xiavg}
\frac{d\xi}{ds}&=&\frac{\Om_K-\om^*}{\eps}-\hat C_K\sin\xi\nonumber\\
&-& \sum_{m=1,m\ne K}^M \hat C_m\sin\left((\Om_m-\Om_K)\frac{s}{\eps} +\Psi_m-\Psi_K+\xi\right).
\end{eqnarray}
The last terms are higher frequency terms since $|\Om_m-\Om_K|\gg \eps$ for $m\ne K$. The method of averaging \cite{ascs} is applied to this equation over $0\le s\le T$. The higher frequency terms average to $O(\eps/(TR^*))$; this estimate holds for each $R^*>0$, but not uniformly in $R^*$. This gives to leading order in $\eps$
\beq\label{eq:barxiavg}
\frac{d\bar\xi}{ds}=\frac{\Om_K-\om^*}{\eps}-\hat C_K\sin\bar\xi, 
\eeq
and the method shows that $\xi(t)-\bar\xi(\eps t)=O(\eps)$ uniformly for $0\le s \le T$. Equation (\ref{eq:barxiavg}) may also be written as a gradient system, and it has an equilibrium $\bar\xi= \arcsin X_K$ where 
$$
X_K(\om^*)=\frac{(\Om_K-\om^*) R^*}{\eps C_K},
$$
if the phase locking condition (\ref{eq:PLcondition}) holds.
The equilibrium is stable under persistent disturbances. 

As a result, for all data  satisfying (\ref{eq:PLcondition}),
$$
\theta=\om^* t+\phi =\om t-\xi+(\Om_K-\om^*) t+\Psi_K \approx \Om_K t +\phi^*_K
$$
where $\phi^*_K=\Psi_K- \arcsin X_K$. It is shown in \cite{cims22} that $\xi = \phi^*_K+O(\exp(-\hat C_K s))$, and so after an initial transient, $\bar\xi$ equilibrates to $\phi^*$.

The phase deviation $\phi$ is not constant, but has frequency $\Om_K-\om^*$. The relation between $\dot\phi$ and $\om^*-\Om_K$ is essentially linear for $\om^*$ near $\Om_K$; this fact facilitates the design of phase locking devices \cite{lindsey}.

\section{Appendix 2: Power Spectrum of $z$.}

The  power spectrum density (PSD) of $f$ given in (\ref{eq:f}) is
$$
PSD_f(\rho)=\sum_{m=1}^M |C_m|^2\delta(\rho-\Om_m).
$$
The approximation to $z$ derived earlier, may be used to approximate the  PSD of $z$ using the formula 
$$
PSD_z(\rho)=|{\cal F}_z(\rho)|^2
$$
where ${\cal F}_z$ is the Fourier transform of $z$ given by
$$
{\cal F}_z(\rho)=\frac\eps{T} \int_{0}^{T/\eps} e^{-\imag\rho t}z(t)\, dt\approx \frac\eps{T} \int_{0}^{T/\eps} e^{-\imag\rho t}R^* e^{\imag\theta(t)}\, dt.
$$
The notation $\om^*\sim\Om_K$ means that the phase locking condition (\ref{eq:PLcondition}) is satisfied, so $\theta\approx \Om_K t +\phi^*_K(\om^*)$; and if $\om^*\sim\Om_K$, then
\begin{eqnarray*}
{\cal F}_z(\rho)&\approx&\frac\eps{T} \int_{0}^{T/\eps} e^{-\imag\rho t}R^* e^{\imag(\Om_K t+\phi^*_K)}\, dt\\
&=&R^*e^{\imag \phi^*_K}\sinc((\Om_K-\rho) T/\eps)\, dt +O(\eps/T).
\end{eqnarray*}
The result is that
$$
PSD_z(\om^*)\approx R^{*2}\,\sinc^2((\Om_K-\om^*) T/\eps)+O(\eps/T).
$$
Calculating the errors made in these approximations is left as an exercise for the reader.

\begin{thebibliography}{99}
\bibitem{buzsaki} G. Buzsaki (2006) Rhythms of the Brain, Oxford U Press.
\bibitem{guttman} R. Guttman, et al. (1980) Frequency Entrainment of Squid Axon Membrane, J Membr Physiol 56:9-18.
\bibitem{nunez} P. L. Nunez, R. Srinivasan (2005) Electric Fields of the Brain: The Neurophysics of EEG, 2nd Ed., Oxford U Press.
\bibitem{emires} E. M. Izhikevich (2001) Resonate-and-Fire Neurons, Neural Networks, 14:883-894.
\bibitem{burst} E. M. Izhikevich,  et al., (2003) Bursts as a unit of neural information: selective communication via resonance. Trends in Neurosci. 26 (3):161-167.
\bibitem{okeefe} J. O'Keefe, M. L. Recce (1993) Phase Relationship Between Hippocampal Place Units and the EEG Theta Rhythm. Hippocampus, 3 (3): 317-330.
\bibitem{singer} P. Fries, D. Nikolić, W. Singer (2007) The Gamma Cycle, Trends in neuroscience, 30 (7): 309-16.
\bibitem{winfree} A. Winfree, (1980), The Geometry of Biological Time, Springer-Verlag, New York.
\bibitem{grol1} J. Torrejon, et al. (2017) Neuromorphic computing with nanoscale spintronic oscillators, Nature, 547:428-431 (27 July 2017)
\bibitem{grol2} M. Romero, et al., (2018) Vowel recognition with four coupled spin-torque nano-oscillators, Nature, v 563 (8 Nov, 2018). (doi.org/10.1038/s41586-018-0632-y).
\bibitem{dynconn} F. Hoppensteadt, E. Izhikevich (1999) Oscillatory Neurocomputers with Dynamic Connectivity, PRL, vol. 82, \#14: 2983-2986.
\bibitem{actpat} F. Hoppensteadt (2009)  Activity patterns in networks stabilized by background oscillations, Biol. Cybernetics 101 (1):43-47.
\bibitem{FCH} F. Hoppensteadt (2015) Spin torque oscillator neuroanalog of von Neumann's microwave computer, BioSystems, 136:99-104
\bibitem{slavin} A. Slavin, V. Tiberkevich  (April 2009) Nonlinear Auto-Oscillator Theory of Microwave Generation by Spin-Polarized Current, IEEE Trans on Magnetics, 45, no. 4:1875.
\bibitem{macia} F. Macia, et al. (2011) Spin-wave interference patterns created by spin-torque nano-oscillators for memory and computation, Nanotechnology 22 (9): 095301.
\bibitem{wcnn} F. Hoppensteadt, E. Izhikevich (2012) Weakly Connected Neural Networks, Springer-Verlag.
\bibitem{cims22} F. Hoppensteadt (2011) Quasi-static State Analysis of Differential, Difference, Integral and Gradient Systems, Courant Lecture Notes 21, Amer. Math Soc.
\bibitem{malkin} I. G. Malkin (1958) Theory of Stability of Motion. AEC Translation Series,. AEC-t-3352.
\bibitem{ascs} F. Hoppensteadt (1991) Analysis and Simulation of Chaotic Systems, 2nd ed., Springer-Verlag, Berlin.
\bibitem{carhop} H. Carrillo, F. Hoppensteadt (2010) Unfolding an Electronic Integrate-and-Fire Circuit, Biol. Cyber, 102:1–8.
\bibitem{flahop} J. E. Flaherty, F. Hoppensteadt (1978). Frequency entrainment of a forced van der Pol oscillator, Studs. Appl. Math., 58(\#1): 5-15.
\bibitem{vcon} F. Hoppensteadt (2006) Voltage-controlled oscillations in neurons, Scholarpedia, 1(11):1599.
\bibitem{wentzel} M.I. Friedlin, A.D. Wentzell (2012) Random Perturbations of Dynamical Systems, Springer-Verlag, Heidelberg.
\bibitem{skor} A. Skorokhod, F. Hoppensteadt, H. Salehi, (2002) Random perturbation methods with applications in science and engineering, Springer-Verlag, New York.
\bibitem{spin2} F. Macia, et al. (2014) Spin wave excitation patterns generated by spin torque oscillators. Nanotechnology 25 (4):045303.
\bibitem{spinwave} F. Macia et al., (2013) Spin wave excitation patterns generated by spin torque oscillators, Arxiv 1309.2436v1.
\bibitem{lindsey} W.C. Lindsey,  C.M. Chie (1986) Phase-locked loops, IEEE, New York.
\end{thebibliography}
\end{document}